# DNN-assisted optical geometric constellation shaped PSK modulation for PAM4-to-QPSK format conversion gateway node

**Takahiro Kodama[1], Toshiaki Koike-Akino[2], David S. Millar[2], Keisuke Kojima[2], Kieran Parsons[2]**
1. Faculty of Engineering and Design, Kagawa University, Takamatsu 761-0396, Japan
2. Mitsubishi Electric Research Laboratories, 201 Broadway, Cambridge, MA 02139, USA

**Abstract:** An optical gateway to convert four-level pulse amplitude modulation to quadrature phase shift keying modulation format having shaping gain was proposed for flexible intensity to phase mapping which exploits non-uniform phase noise. The power consumption of the optical modulation format conversion can save by making a DNN-based decision on the receiver side for the generated QPSK signal with non-uniform phase noise. A proof-of-principle experiment has shown that an optically geometric constellation shaped QPSK modulated signals generated from regular PAM4 signals with Gaussian-distributed noise. The shaped QPSK signal shows BER and generalized mutual information improvement by 1dB gain through the use of digital neural network signal recovery.

**Keywords:** Optical modulation format conversion, machine learning, digital coherent communication, optical signal processing.

## 1. Introduction

In the current optical communication network (NW), the different types of modulation formats and detection schemes have been used; specifically, intensity modulation/direct detection (IM/DD) and phase modulation/coherent detection, respectively, for short-reach and long-reach fiber NWs. Indirect modulation format conversions can cause a traffic delay due to optical-electrical-optical (O/E/O) conversion at the heterogeneous NW gateway node (HNGN) connecting short- and long-reach optical fiber communication NWs [1-4].

Two important technologies for transmitting from a short-distance NW to a long-distance NW without O/E/O conversion are IM/coherent detection at the transmitter/receiver and conversion from pulse-amplitude modulated (PAM) signal to the phase-shift keying (PSK) signal gateway node. Direct modulation format and wavelength conversion in the optical-domain converting from a PAM to PSK signal to an IQ modulated signal using nonlinear optical effects such as cross-phase modulation (XPM) and four-wave mixing (FWM) has been demonstrated to construct a future optical-based HNGN [5-9]. Research on PSK to PAM signal conversion [10-11] and PSK to quadrature amplitude modulation conversion [12-15] using similar nonlinear optical effects has been demonstrated for a long period. When generating nonlinear optical effects, semiconductor optical amplifier such as active device is superior in miniaturization. Still, it has limitations in speeding up, so passive devices such as HNLF are more advantageous. However, since the nonlinear coefficient is small, there is a problem that the fiber length is required and the device size becomes large. A definite advantage of optical modulation format conversion is that it seamlessly transmits from short-reach optical NW to a long-reach optical NW while keeping the packet format, modulation depth, and symbol rate without using a high-speed driver amplifier and linear optical IQ modulator [16].

Recently, optical-domain PAM4 to QPSK modulation format conversions have been demonstrated [17]. Although the IM signal was assumed to have a high signal-to-noise ratio (SNR) in the previous work, there is a situation where the intensity-modulated signal has a low SNR due to the poor characteristics of the light source and the optical amplifiers. When converting from a PAM signal with low SNR to PSK signal, the following two phenomena occur. (1) converting from intensity noise to phase noise, (2) large phase noise as increasing the amount of phase change. Especially, PAM4 to QPSK modulation format conversion needs to account for the amount of four-level phase change of QPSK signal in consideration of the noise amount of the PAM signal to improve the signal quality.

In the present paper, we propose a flexible generation of optical-domain geometric-constellation shaped (OGS) QPSK signal using the optical amplifier level tuning of regular PAM4 signal, and an optimum decision of OGS-QPSK signal using coherent detection and deep neural network (DNN) in the digital signal processing (DSP). To confirm the feasibility of the proposed hybrid processing of optical modulation format conversion and DNN at the first stage, we experimentally proofed the generation of non-uniform phase noise distributed QPSK signal from Gaussian noise-added PAM4 using XPM generated in high nonlinear fiber (HNLF). We show that the DNN-based decision can significantly outperform a conventional hard-decision and that the penalty due to the non-optimal amplifier level can be effectively compensated by DNN for variable OGS-QPSK signal.

## 2. Operation principle

Figure 1 shows the configuration of the proposed PAM4-to-QPSK modulation format conversion using OGS and DNN-based decision. At the transmitter (Tx) side, non-return-to-zero (NRZ) PAM4 generated by laser diode (LD) or Lithium niobite intensity modulator (LN-IM) at a wavelength $\lambda_1$ from the short-reach optical communication NW

At the HNGN, the optical switch selects whether to pass the PAM4 signal through the long-reach optical communication NW or drop it once with HNGN. When dropping, it is received by PD in the same way as standard PAM4 signal. When passing, the PAM4 signal and continuous wave (CW) as a probe light at wavelength $\lambda_2$ are launched into an HNLF. The probe light is modulated in its phase due to XPM induced by the PAM4 with tuning optical amplifier level at the erbium doped fiber amplifier (EDFA). The amount of phase change of the probe light due to XPM can be expressed as

$$\Delta\varphi(k) \cong k\left(2\gamma_1 L_{eff} P_k\right), \tag{1}$$

where $\gamma_1$ and $L_{eff}$ are the nonlinear coefficient and the effective interaction length of the HNLF; $P_k$ is the peak power of $k$-th level of a PAM4 signal. The absolute value of the electric field amplitude of PAM4 this time is arranged at unequal intervals in the same as the electric field amplitude $E_k$ of standard PAM4 in consideration of square detection by a photodetector. The electric field amplitude at each level can be expressed as $E_k \propto \sqrt{P_k}$. In the case of high SNR of PAM4 signal, the almost ideal QPSK signal with $\Delta\varphi(k) = \frac{\pi}{2}k$ is generated, as shown in Fig. 2(a). Our previous studies have evaluated SMF transmission's effect after modulation format conversion for high SNR PAM4 signals and suggested the possibility of low power consumption operation when the HNLF input power is reduced from the optimum power [17].

As of decreasing SNR of the PAM4 signal, the QPSK signal with $\Delta\varphi(3) = \frac{3\pi}{2}$ is drastically degraded due to phase noise, as shown in Fig. 2(b). Figure 2(c) shows the OGS-QPSK signal that can mitigate phase noise by tuning the optical amplifier level. It is confirmed that the phase difference between adjacent signal points is less than π/2. If $\lambda_1$ and $\lambda_2$ are too close, there is a low crosstalk tolerance for the PAM4 signal; if we increase the wavelength detuning, the XPM effect becomes small due to the walk-off between two wavelengths. The parameters of the HNLF, nonlinear coefficient, length, chromatic dispersion, and so on, shall be chosen carefully since those may lead to undesired FWM and unstable phase modulation in the PAM4-QPSK conversion. An optical bandpass filter (OBPF) is placed after the HNLF to divide the generated QPSK signal from the PAM4 signal. At the end node of long-reach optical communication NW, the QPSK signal is detected by a coherent detection with high receiver sensitivity, and processed by DNN-based decision.

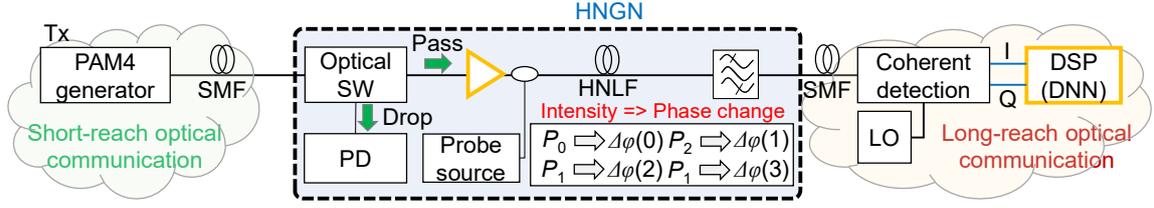

Fig. 1. OGS-QPSK generation and DNN decision.

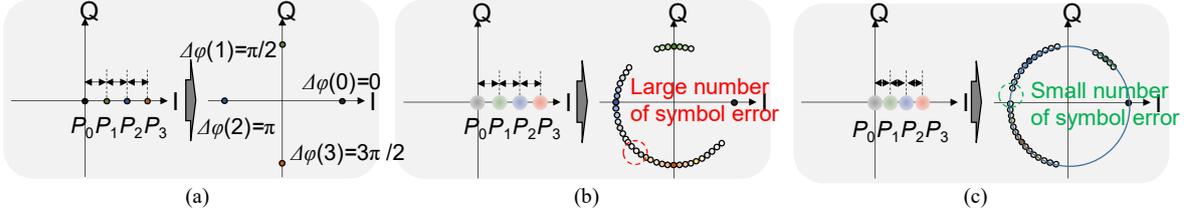

Fig. 2. Schematic diagram of PAM4-to-QPSK format conversion, (a) regular QPSK for noiseless PAM4, (b) regular QPSK by PAM4 at low SNR, (c) OGS-QPSK by PAM4 at low SNR.

## 3. Experiment of non-uniform phase noise distributed QPSK generation

Figure 3 shows the experimental setup for the proof-of-demonstration of OGS-QPSK generation from PAM4 signal. This is achieved by generating a non-linear optical effect that is smaller than the standard QPSK signal and making the phase difference between adjacent signal points less than $\pi/2$. Here, we focus only on the nonlinear optical effects generated by HNGN. We do not consider the nonlinear optical effects generated when the long-distance transmission is performed over access NWs and metro networks.

At the transmitter (Tx) side of an edge node in the access NW, a narrow line-width tunable LD with a line-width 100 kHz at the central wavelength of 1550 nm generated the unmodulated lights. The output of the LD was connected to a LN-IM. Here, we used a narrow line-width LD at 1550 nm for both a probe light source in the HNGN and a local oscillator in the receiver. At the electrical domain, the Tx side of ONU in access NW, a 10 Gbaud NRZ-PAM4 with a pseudo random bit sequence (PRBS) data pattern with 10 Gsample/s was generated by an arbitrary waveform generator (AWG: Tektronix, 7122C) and driver amplifier, and an optical NRZ-PAM4 signal was obtained by modulating the CW light at 1550 nm by the LN-IM. The low SNR PAM4 signal was emulated by reducing the gain of the driver amplifier. The clock signal output from the AWG is used to synchronize the AWG and digital storage oscilloscope (DSO: Tektronix, 6154C).

At the HNGN, the optical PAM4 signal and 2.4 dBm average power of the probe light at 1545 nm were launched into the HNLFs. The parameters of the HNLFs were summarized in Table 1. The overall loss of the HNLF including propagation loss and connection loss was 3.2 dB. The amounts of noise and phase change due to XPM in the HNLF were controlled by a gain of an EDFA before the HNLFs. The polarization controllers were arranged and adjusted so that the two input lights' polarization states match in the HNLFs. After HNLF, the optical filter with a transmission band of 1 nm passes only the QPSK signal generated at the center wavelength of 1545 nm.

At the receiver (Rx) side of an edge node in the metro NW, the received OGS-QPSK signal adjusted the average power by a variable optical attenuator (VOA) and mixed with the local oscillator light, and coherently detected. The received signal was sampled and quantized by an analog to digital converter in a DSO. The output signal with 20 Gsample/s of the DSO was processed offline such as down sampling, frequency offset compensation (FOC) and decision directed-based carrier phase recovery (CPR). Figures 3(b, c) show the constellation and phase distribution of the received OGS-QPSK signal with the different phase noise for four-level. From the result of Fig. 3 (c), we can confirm that the phase noise amount increases as the phase change amount increases. The signal distribution of the signal points without phase change due to XPM becomes shot noise generated in the receiver.

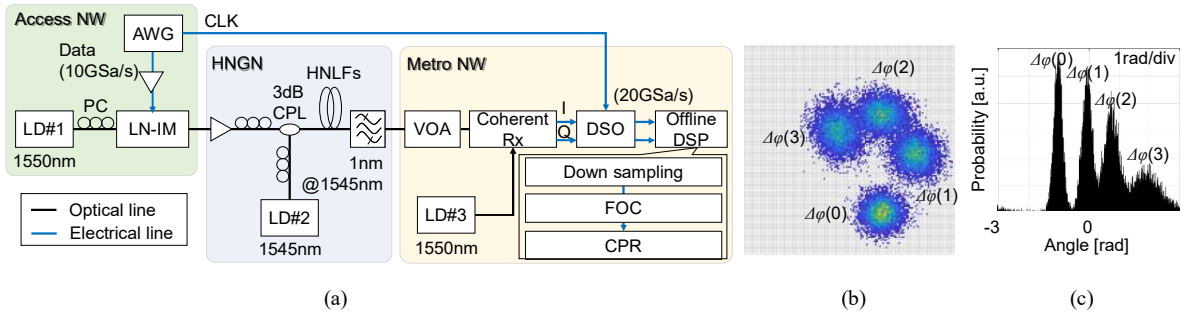

Fig. 3. Experimental results, (a) setup, (b, c) received constellation and phase distribution diagram.

Table 1. HNLF parameters

| HNLF parameters on 1550nm | #A | #B |
|---|---|---|
| Length [km] | 2.5 | 1 |
| Nonlinearity [1/W/km] | 10 | 10 |
| Dispersion [ps/nm/km] | 0.57 | 0.52 |
| Slope [ps/nm/km$^2$] | 0.018 | 0.016 |
| Loss [dB/km] | 1.07 | 0.76 |

## 4. Simulation setup and results

To verify the performance of the 10 Gbaud XPM-based NRZ-PAM4-to-QPSK modulation format conversion using OGS and DNN-based decision, we conducted a numerical simulation for evaluating the effectiveness of the proposed method. We evaluated an additive white Gaussian noise (AWGN) tolerance of the PAM4 signal by changing the geometric constellation shape. The parameters of the HNLFs were same condition as experiment and the connection loss between HNLFs was assumed to be zero. The central wavelengths for both a probe light and a signal light were set to 1540 nm and 1550 nm, respectively. The walk-off effect can be ignored because a dispersion slope value is small. In this evaluation, AWGN adding after modulation format conversion was not considered for simplicity of analysis. Figures 4 show constellation maps of the OGS-QPSK signal at an SNR of PAM4 signal from 20 dB to 30 dB, and an HNLF input power from 44 mW to 60.5 mW. When the HNLF input power is 55 mW, the standard QPSK signal with $\Delta\varphi(k) = \frac{\pi}{2}k$ is generated in the absence of PAM4 noise. However, noise at PAM4 causes a considerable distortion of QPSK signal especially at $\Delta\varphi(3) = \frac{3\pi}{2}$. We aim at designing the phase shift of QPSK constellations to exploit the non-uniform phase distortion characteristic caused by the optical gateway.

To compensate for such irregular distortion of shaped QPSK constellations, we employ DNN whose architecture is depicted in Fig. 5, where multiple fully-connected linear layers, batch normalization layers, rectified linear unit (ReLU) activation layers, and skip connection are configured with 10% dropouts. The DNN feeds distorted QPSK symbols to generate 2-bit log-likelihood ratios (LLRs), whose sigmoid cross-entropy loss is minimized by a stochastic gradient descent based on adaptive momentum. The data patterns for the DNN learning and testing were 120,600 bits and 13,400 bits, respectively.

Figure 6(a) shows the bit error ratio (BER) performance of the OGS-QPSK signal as a function of SNR for the PAM4 signal. In the case of the conventional linear equalization, the best BER is obtained when the HNLF input power is 55 mW. When the HNLF input power is reduced to 38.5 mW, a terrible BER performance is seen regardless of the SNR range. Whereas, in the case of DNN-based decision, the BER degradation can be fully compensated and the BER performance at the HNLF input power of 38.5 mW is slightly improved compared to the standard 55 mW case. This is a benefit of DNN decision and optical constellation shaping, which controls phase shifts according to non-uniform phase distortion.

As the DNN-based signal recovery produces soft-decision LLR values, the shaping gain is more remarkable in terms of the generalized mutual information (GMI) against the traditional BER metric. Figure 6(b) shows the GMI performance of the OGS-QPSK signal as a function of PAM4 SNR. When the optical HNLF input power of 38.5 mW, the achievable gain at a GMI of 0.8 will be nearly 1 dB compared to non-shaped QPSK phase shifts with 55 mW input power. In addition, the DNN signal recovery has an outstanding performance improvement by greater than 3 dB over the conventional linear equalization. In consequence, the proposed OGS and DNN-based decision can save the required HNLF input power.

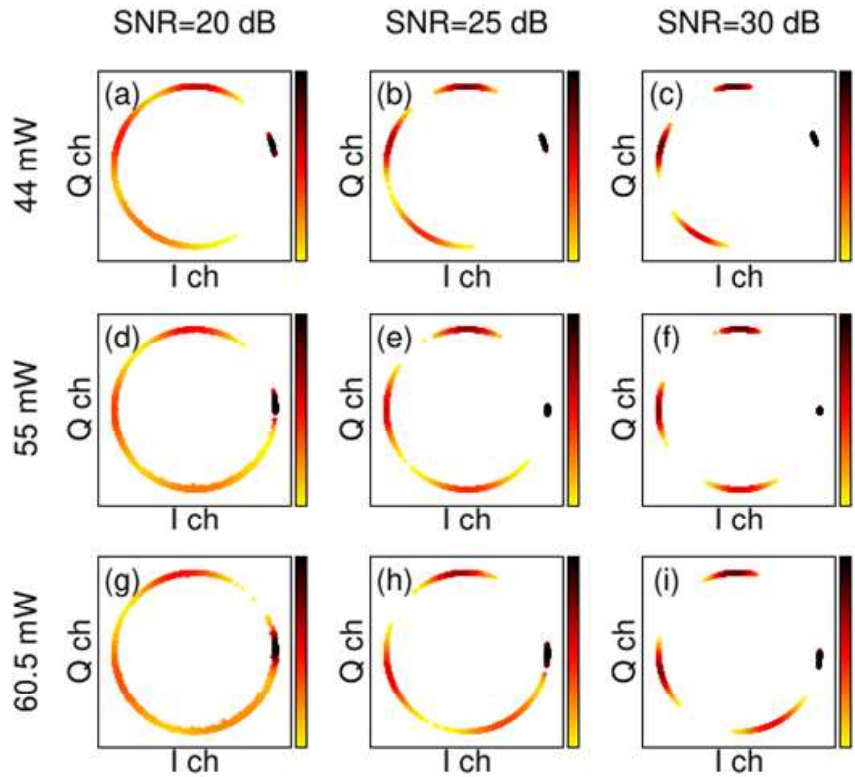

Fig. 4. Constellation maps of OGS-QPSK sign.

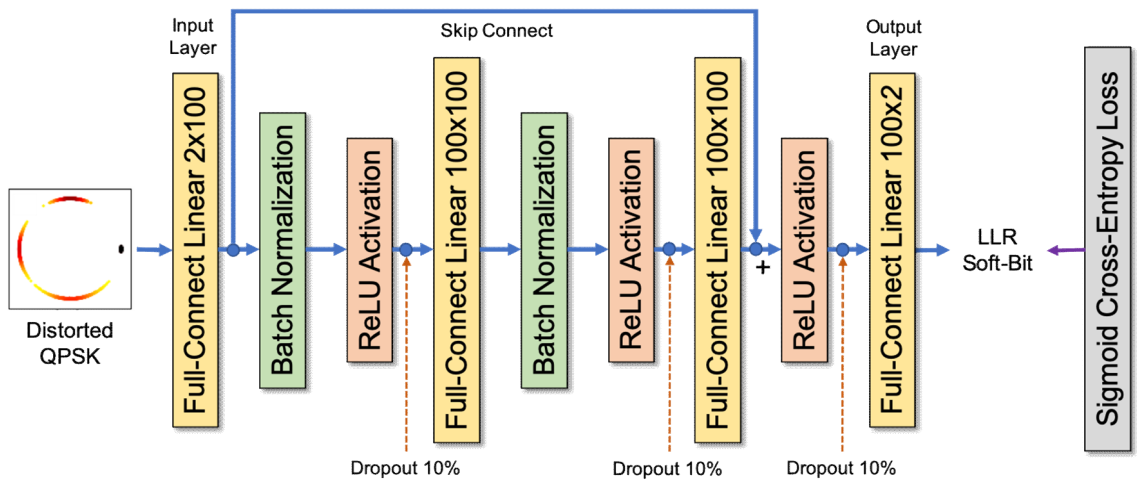

Fig. 5. DNN-based signal recovery.

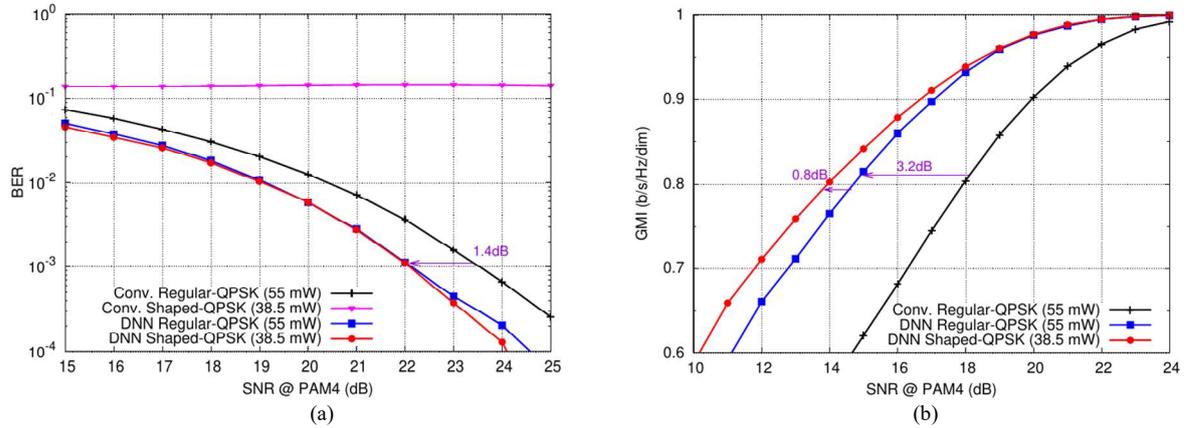

Fig. 6. Performance of OGS-QPSK as a function of SNR for PAM4 in HNGN modulation converter. (a) BER and (b) GMI

## 5. Conclusions

We proposed and experimentally demonstrated an OGS-QPSK generation for optical PAM4-to-QPSK conversion at HNGN. The 10 Gbaud OGS-QPSK signal with DNN-based decision can reduce the HNLF input power of the PAM4 signal compared to the original QPSK with the conventional linear equalization. Specifically, the DNN-based signal recovery achieves approximately 3 dB gain, and another gain of nearly 1 dB can be achieved by optical constellation shaping. To the best of authors' knowledge, there is no other literature which applied deep learning techniques to mitigate nonlinear distortion occurred in the HNGN.